# Russel and Rao Coefficient is a Suitable Substitute for Dice Coefficient in Studying Restriction Mapped Genetic Distances of *Escherichia coli*

Zhu En Chay[1], Chin How Lee[1], Kun Cheng Lee[1],
Jack SH Oon[1], Maurice HT Ling[1,2]

[1]*School of Chemical and Life Sciences*
*Singapore Polytechnic, Singapore*

[2]*Department of Zoology*
*The University of Melbourne, Australia*

**Abstract**

*Escherichia coli* is one of many bacterial inhabitants found in human intestines and any adaptation as a result of mutations may affect its host. A commonly used technique employed to study these mutations is Restriction Fragment Length Polymorphism (RFLP) and is proceeded with a suitable distance coefficient to quantify genetic differences between 2 samples. Dice is considered a suitable distance coefficient in RFLP analyses, while others were left unstudied in its suitability for use. Hence, this study aims to identify substitutes for Dice. Experimental data was obtained by subculturing *E. coli* for 72 passages in 8 different adaptation media and RFLP profiles analyzed using 20 distance coefficients. Our results suggest that Dennis, Fossum, Matching and Russel and Rao to work as well or better than Dice. Dennis, Matching and Fossum coefficients had highest discriminatory abilities but are limited by the lack of upper or lower boundaries. Russel and Rao coefficient is highly correlated with Dice coefficient ($r^2 = 0.998$), with both higher and lower boundaries, suggesting that Russel and Rao coefficient can be used to substitute Dice coefficient in studying genetic distances in *E. coli*.

## 1   Introduction

*Escherichia coli* is a Gram-negative bacteria species that inhabits in the gastrointestinal tract of humans (Foley et al., 2009) and is one of the most thoroughly studied organism (Welch et al., 2002). It is a diverse species where some *E. coli* strains live as harmless bacterium (Welch et al., 2002), while other strains like O157:H7 can cause a wide range of intestinal and extraintestinal diseases (MacDonald et al., 1988; Clermont et al., 2007). *E. coli* has been identified as one of the major bacterial foodborne infections and due to their significant impact on human health (MacDonald et al., 1988), many molecular methods include restriction endonuclease analysis, polymerase chain reaction and DNA sequence polymorphism were derived to study them (Foley et al., 2009). *E. coli* may be genetically altered by diets of its human host as it had been suggested that *E. coli* has higher prevalence to antibiotics resistance (Silva et al., 2007). Oral antibiotics consumption caused emergence of antibiotic resistant strains of *E. coli* (Bourque et al., 1980) and infections caused by antimicrobial-resistant bacteria are associated with substantial morbidity and mortality (O`Fallon et al., 2009).

Nucleic acid fingerprinting are analysis techniques employed to differentiate between DNA samples based on its DNA band patterns generated after enzymatic amplification of variable regions and analyzed by gel electrophoresis, with or without restriction digestion (Gilbride et al., 2006). The project will utilize RFLP method to study the DNA bands.

A distance coefficient is to quantify comparable features (similarity and differences) of two given vectors between two objects where the collective differences or dissimilarity can be denoted as a distance measure, seen as a scalar measure (Basilevsky, 1983). There are different ways in computing distance coefficients, each differing from each other as the emphasis is given to either the intersecting area or the non-intersecting regions may differ. Most distance coefficients have upper and lower boundaries, distinct to the mathematical equation used. If defined, the lower boundary of a coefficient denotes complete difference, whereby the upper boundary suggests complete similarity. Values within the 2 boundaries are scaled to determine similarities and differences The values given by each distance coefficient varies (Duarte et al., 1999). Thirty-five of these distance coefficients were compiled (Ling, 2010).

Genetic distance refers to the genetic difference and similarity between and within a species that is often used for classification and evolutionary studies involving humans, mammals, fruit flies and mosquitoes with the involvement of statistical models (Wang et al., 2001). A smaller genetic distance indicates a closer genetic relationship, while a larger genetic distance will indicate a weaker genetic relationship in comparison studies, useful for reconstructing phylogenetic relationships (Shriver et al., 1995). The commonly discussed genetic distance measures are Nei's minimum genetic distance and Nei's standard genetic distance. The two genetic distance measures are nonlinear with time or large mutation rates considered as factors. The linearity of Nei Li genetic distances is factored by the frequency of mutations (Nei & Li, 1979; Shriver et al., 1995). Some other proposed genetic distance measures include average square distance (ASD) (Goldstein et al., 1995a), Delta Mu Genetic Distance (Goldestein et al., 1995b), stepwise weighted genetic distance measure (*Dsw*) (Shriver et al., 1995), kinship coefficient (*Dkf*) (Cavalli-Sforza & Bodmer, 1971) and coancestry coefficient *Theta* (*Fst*) (Reynolds et al., 1983).

Russel and Rao is a distance measure used for dichotomous variables (Hwang et al., 2001; Russel & Rao, 1940). It was previously studied for random amplified polymorphic DNA (RAPD) and was concluded that Russel and Rao can only be used for specific instances due to its exclusion of negative co-occurrences (Coefficient D) in numerator and inclusion in the denominator (Meyer et al., 2004).

## 2    Objectives

This study aims to determine suitable distance coefficient measures from 20 of the 35 compiled measures (Ling, 2010) to study the genetic distance, at the genomic scale, of a sequenced strain of human intestinal bacterium, *Escherichia coli* ATCC 8739. Since Dice is the only ideal distance coefficient studied for its use in RFLP *E. coli* genetic study analysis, this study aims to identify other distance coefficients that are capable of substituting Dice. Restriction Fragment Length Polymorphsm (RFLP) is employed in this study for being economical, fast and simple compared to other DNA fingerprinting methods (Xiao et al., 2006). The band patterns produced as a result of varying lengths of restriction fragments will be analyzed using statistical methods (Nei & Li, 1979). Measurement errors tend to occur and hide differences between different RFLP studies and statistical methods are used to normalize these errors (Evett et al., 1993). Our results demonstrated that Russel and Rao coefficient can be used as substitute for Dice coefficient in studying genetic distances in *E. coli*.

## 3 Methodology

**Bacterial culture and PCR-RFLP DNA Fingerprinting.** *Escherichia coli* ATCC 8739 was inoculated into 8 different treatment supplementation in Nutrient Broth [0.025% (w/v) as high MSG (H MSG); 0.0025% (w/v) as low MSG (L MSG); 0.025% (w/v) as high benzoic acid (H BA); 0.0025% (w/v) as low benzoic acid (L BA); 1% (w/v) NaCl as high salt (H SALT), Nutrient Broth as low salt (L SALT); H MSG, H BA and H SALT as high combination (H COMB); L MSG and L BA as low combination (L COMB)] and cultured at $37^oC$. Subculture was performed every 2 to 3 days from 1% of the previous culture. Genomic DNA was extracted from the treatment cultures at every $12^{th}$ passages for Polymerase Chain Reaction (PCR) and Restriction Fragment Length Polymorphism (RFLP). A total of 72 subcultures were carried out, resulting in 6 time-points. A total volume of 50μl in each PCR reaction was prepared according to supplier's specification (New England Biolabs, Inc.), consisting of 1 of the 3 primers: Primer 5, CgCgCTggC; Primer 6, gCTggCggC; and Primer 7, CAggCggCg. Each of the primers act as both forward and reverse primers. The PCR reaction was carried out (Hybaid Limited, PCR Express) with the cycling condition of initial denaturation at 95˚C for 10 minutes; 35 cycles of amplification at 95˚C for 1 minute, 27˚C for 1 minute, 72˚C for 3 minutes; followed by a final extension at 72˚C for 10 minutes. The product was digested with a unit of restriction endonuclease (TaqI, HinfI or MspI) for 16 hours. Both PCR and RFLP products were visualized on 2% (w/v) agarose gel with 1X GelRed. A total of 12 agarose gels per time-point (3 PCR gels and 3 RFLP gels for each PCR gel) resulting in a total of 72 agarose gels for the 6 time-points under study.

**Determination of Suitable Distance Coefficients.** The experimental data obtained from RFLP will have the bands of a lane measured and retention factor (Rf) value are obtained. Lanes in a gel will be compared with each other using 20 distance coefficients (Ling, 2010). Results obtained from the 20 distance coefficient measures are then compared with other distance measures using analysis of variance (ANOVA), coefficient of determination ($r^2$) and the arithmetic means, standard deviation (SD) and Coefficient of Variance (COV) at percentiles ranges 0 - 10, 10 – 90 and 90 – 100 for the identification of suitable distance coefficients for use of RFLP analysis of *E. coli*. There will be a total of 190 pair-wise combinations arises from 20 permutations of 2, possible from the 20 distance coefficients. As there are 72 gels of data, 190 pair-wise combinations will be done for every gel, giving 13680 possible number of comparisons to be done. Each gel has 8 different treatments, comparison of either 2 treatments gives 28 different combinations. Each distance coefficient will have 2016 number of analysis to be done, derived from 28 number of combinations of 72 gels. However, some comparisons are excluded if there are no observable bands, such as DNA smear. A total of 383,040 (28 x 13680) number of comparison studies to be done. The suitable distance coefficients should cover a broad spectrum of data that is capable of comprising extremely huge and small data values, with ability to discriminate small data changes and differences.

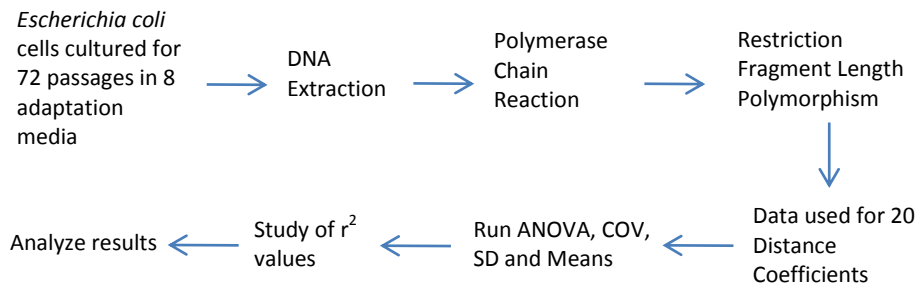

**Figure 1:** Flowchart of Methodology used in this study

## 4 Results and Discussion

A one-way ANOVA test was run to first detect differences in the results of the 20 distance coefficients' (Table 1). A p – value of less than 0.001 demonstrate statistically significant differences in the 20 distance coefficients, suggesting that the distance coefficients' result outcomes are incomparable. To study the incomparability, the analysis of COV values in the 3 percentile ranges are done to identify the ideal distance coefficient and $r^2$ value for each test is obtained to study similarities.

| Source | Sum of Squares | D. f | Mean Square | F ratio | P - value |
|---|---|---|---|---|---|
| Between Groups | 55,205.4 | 19 | 2905.55 | 4405.23 | <0.001 |
| Within Groups | 23,594.2 | 36318 | 0.659567 | | |
| Total | 79,159.6 | 36337 | | | |

**Table 1:** Results of ANOVA of the 20 Distance Coefficients.

The mean, SD and COV of the 20 distance coefficients in Table 2 are arranged in an descending order of the number of percentile range wins against Dice. The suitable distance coefficient can be deduced by observing the coefficient of variance (COV) at the 3 range of percentiles. A high COV at 0 – 10 percentile range confers to a distance coefficient with high capability to detect and discriminate low values of genetic distances, while a low COV be otherwise. Meanwhile a high COV at 90 – 100 percentile range illustrates a distance coefficient with good detection and distinction capacity for high values, a low COV be otherwise. For a distance coefficient with a high COV at 10 – 90 percentile range demonstrates high capability to cover the majority of values with acceptable discriminative ability, while a low COV be otherwise.

Dice distance coefficient $Dice = \frac{2A}{(A+B)+(A+C)}$ [Dice, 1945] is identical to Nei Li distance coefficient $Nei\ Li = \frac{2n_{xy}}{n_x+n_y}$ (where $n_x$ and $n_y$ are number of fragments in populations X and Y respectively; whereas $n_{xy}$ is the number of fragments shared by two populations) (Nei & Li, 1979] and Nei Li was tested to be a reliable distance coefficient from a simulation study (Li, 1981). Since Nei Li (Nei & Li, 1979) has been identified as a reliable distance coefficient, it will be used to identify suitable distance coefficients. Coefficients B, C and D in Dice are substituted with 0 and yield a upper boundary of 1, while A and D gave a lower boundary of 0. COV values of Dice (Dice, 1945) will be used as a threshold to identify distance coefficients that are similar or better than Nei Li. Dice (Dice, 1945) will be used to represent Nei Li (Nei & Li, 1979) in this study. With this, the benchmark for percentile range 0 – 10, 10 – 90 and 90 – 100 are 0.286, 0.297 and 0.000 respectively.

Forbes (Forbes, 1907), Anderberg (Ling, 2010), Sokal and Sneath (Sokal & Sneath, 1963), Hamann (Hermann, 1961), Roger and Tanimoto (Roger & Tanimoto, 1960), McConnaughey (McConnaughey, 1964), Jaccard (Jaccard, 1908), Sokal and Michener (Sokal & Michener, 1985), Gower and Legendre (Gower & Legendre, 1986), Tulloss (Tulloss, 1997), Buser (Holliday et al., 2002), Sokal and Sneath 2 (Sokal & Sneath, 1963), Ochiai (Ochiai, 1957), Kulczynski 2 (Holliday et al., 2002) and Simpson (Fallaw, 1979) will not be efficient for use as distance coefficient in studying RFLP of *E. coli* (Table 2), since its COV values do not surpass Dice at 3 percentile ranges. The COV study suggests the above mentioned 15 distance coefficients have poorer discriminatory abilities than Dice (Dice, 1945). A suitable distance coefficient should possess higher COV values in all 3 ranges of percentiles than Dice (Dice, 1945). Dennis (Dennis, 1965), Matching (Dunn & Everitt, 1982), Fossum (Fossum, 1966) and Russel and Rao (Russel & Rao, 1940) were identified to work as well or better than Dice (Dice, 1945) (Table 2). The upper and lower boundaries of Dennis (Dennis, 1965), Matching (Dunn & Everitt, 1982), Fossum (Fossum, 1966) and Russel and Rao (Ruseel & Rao, 1940) were studied.

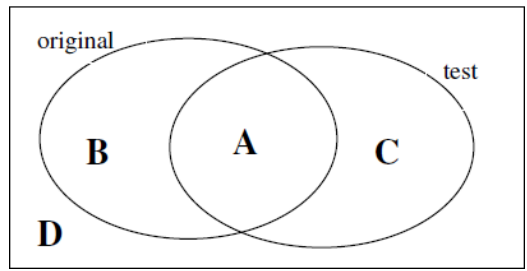

**Figure 2:** Venn diagram illustrating the overlapping regions between two objects. 'A' is the region of intersection, 'D' signifies elements not present in 'original' and 'test' while 'B' and 'C' are elements present in original and test respectively (Ling, 2010).

**Dennis Coefficient**

$$Dennis\ coefficient = \frac{(A \times D) - (B \times C)}{\sqrt{(A + B + C + D)(A + B)(A + C)}}$$

The Dennis (Dennis, 1965) equation coefficients *B*, *C* and *D* that will be replaced with the value 0 when 2 data sets share exact similar numbers, resulting in the upper boundary of 0. However, in the scenario of 2 datasets which share no similar data, a different conclusion of no lower boundary is observed. A relationship was observed, the Dennis coefficient decreases in value with an increasing number of data in 2 datasets, due to coefficient B and C and the negative sign in the quotient.

**Matching Coefficient**

$$Matching\ coefficient = \frac{A + D}{(A + B) + (A + C)}$$

The Matching (Dunn & Everitt, 1982) equation has coefficients B, C and that will be replaced with value 0 when 2 data sets share similar numbers. There is no upper boundary for Matching (Dunn & Everitt, 1982). When A and D coefficient of Matching (Dunn & Everitt, 1982) is replaced with value 0, the result is 0, which is the lower boundary. Hence, Matching (Dunn & Everitt, 1982) coefficient no upper boundary and a lower boundary of 0.

**Fossum Coefficient**

$$Fossum\ coefficient = \frac{(A + B + C + D)(A - 0.5)^2}{(A + B)(A + C)}$$

The Fossum (Fossum, 1966) equation has coefficients B, C and D that will be replaced with the value 0 when 2 data sets share exact similar numbers. There is no upper boundary in Fossum (Fossum, 1966). For the scenario of 2 datasets sharing no similar data, a different conclusion of a lower boundary of 0 is observed. The Fossum coefficient (Fossum, 1966) decreases in value with increasing number of data in 2 datasets, due to coefficient B and C in the quotient and denominator.

**Russel Rao Coefficient**

$$Russel\ Rao\ coefficient = \frac{A}{A + B + C + D}$$

Russel and Rao (Russel & Rao, 1940) has coefficients B, C and D to be replaced with the value 0 when 2 data sets share exact similar numbers. This will give a upper boundary of 1. When coefficients A and D are substituted with value 0 to find the lower boundary, Russel and Rao distance coefficient (Russel & Rao, 1940) yields 0. Russel and Rao (Russel & Rao, 1940) has an upper boundary of 1 and a lower boundary of 0.

With the absence of a lower boundary in Dennis (Dennis, 1965) and Matching (Dunn & Everitt, 1982), and upper boundary in Fossum (Fossum, 1966), the results of the three distance coefficients are hard to interpret. As Dice (Dice, 1945) have a lower and upper boundary, values within the 2 boundaries are scaled to determine similarities and differences where as Dennis (Dennis, 1965), Matching (Dunn & Everitt, 1982) and Fossum (Fossum, 1966) do not have the boundaries to do so. This suggests that both Dennis (Dennis, 1965) Matching (Dunn & Everitt, 1982) and Fossum (Fossum, 1966) are not ideal for use as a distance measure. Meanwhile Russel and Rao (Russel & Rao, 1940) has an upper boundary and lower boundary, suggesting that it can replace Dice (Dice, 1945).

| Distance Coefficient | 0-10 Percentile range | | | 10-90 Percentile range | | | 90-100 Percentile range | | |
|---|---|---|---|---|---|---|---|---|---|
| | Mean | (SD) | COV | Mean | (SD) | COV | Mean | (SD) | COV |
| Dennis | -0.123 | 0.222 | 1.798 | 0.143 | 0.435 | 3.036 | 1.490 | 0.435 | 0.326 |
| Matching | 0.244 | 0.073 | 0.301 | 0.490 | 0.290 | 0.593 | 1.126 | 0.601 | 0.534 |
| Fossum | 0.741 | 0.422 | 0.569 | 6.300 | 3.012 | 0.478 | 12.287 | 1.068 | 0.087 |
| Dice | 0.181 | 0.052 | 0.286 | 0.713 | 0.212 | 0.297 | 1.000 | 0.000 | 0.000 |
| Russel and Rao | 0.182 | 0.052 | 0.285 | 0.714 | 0.212 | 0.296 | 1.000 | 0.000 | 0.000 |
| Forbes | 0.938 | 0.117 | 0.125 | 1.071 | 0.544 | 0.508 | 2.224 | 1.174 | 0.528 |
| Anderberg | 0.144 | 0.058 | 0.405 | 0.697 | 0.245 | 0.352 | 1.000 | 0.000 | 0.000 |
| Sokal and Sneath | 0.144 | 0.058 | 0.405 | 0.697 | 0.245 | 0.352 | 1.000 | 0.000 | 0.000 |
| Hamann | -0.382 | 0.220 | 0.577 | 0.646 | 0.309 | 0.478 | 1.000 | 0.000 | 0.000 |
| Roger and Tanimoto | 0.188 | 0.076 | 0.407 | 0.727 | 0.220 | 0.303 | 1.000 | 0.000 | 0.000 |
| McConnaughey | 0.160 | 0.191 | 1.195 | 0.789 | 0.188 | 0.239 | 1.000 | 0.000 | 0.000 |
| Jaccard | 0.248 | 0.090 | 0.363 | 0.796 | 0.180 | 0.225 | 1.000 | 0.000 | 0.000 |
| Sokal and Michener | 0.309 | 0.110 | 0.356 | 0.823 | 0.154 | 0.188 | 1.000 | 0.000 | 0.000 |
| Gower and Legendre | 0.331 | 0.104 | 0.316 | 0.845 | 0.142 | 0.168 | 1.000 | 0.000 | 0.000 |
| Tulloss | 0.838 | 0.170 | 0.202 | 0.804 | 0.214 | 0.266 | 0.822 | 0.227 | 0.276 |
| Buser | 0.382 | 0.102 | 0.268 | 0.846 | 0.135 | 0.160 | 1.000 | 0.000 | 0.000 |
| Sokal and Sneath 2 | 0.461 | 0.133 | 0.289 | 0.895 | 0.098 | 0.109 | 1.000 | 0.000 | 0.000 |
| Ochiai | 0.476 | 0.098 | 0.206 | 0.884 | 0.107 | 0.121 | 1.000 | 0.000 | 0.000 |
| Kulczynski 2 | 0.580 | 0.096 | 0.165 | 0.894 | 0.094 | 0.105 | 1.000 | 0.000 | 0.000 |
| Simpson | 0.761 | 0.141 | 0.185 | 0.998 | 0.011 | 0.011 | 1.000 | 0.000 | 0.000 |

**Table 2:** The Mean, Standard Deviation and Coefficient of Variance of the 20 Distance Coefficients. The 20 Distance Coefficients are arranged in a descending order of the number of percentile range wins against Dice. E.g: Fossum has COV values at all 3 percentile range higher than Dice, hence arranged above Dice.

Based on Pearson correlation data, Matching (Dunn & Everitt, 1982), Forbes (Forbes, 1907) and Dennis (Dennis, 1965) are inversely proportional to Dice, with their Perarson correlation values to be -0.175, -0.120 and -0.100 respectively. A negative value in Pearson correlation will mean that with increasing Dice coefficient result will yield decreasing Matching, Forbes and Dennis coefficient results. This will make results difficult to interpret; hence, unable to be used as substitute for Dice.

The Coefficient of determination ($r^2$) values of the 20 distance coefficients were obtained from comparing data values. This was to identify correlation relationships between distance coefficients. Figure 2 shows that there are no distance coefficients that are statistically similar to either Dennis (Dennis, 1965), Matching (Dunn & Everitt, 1982) or Fossum (Fossum, 1966). Hamann (Hamann, 1961) is directly correlated to Sokal and Michener (Sokal & Michener, 1958), Sokal and Sneath (Sokal & Sneath, 1963) to Anderberg (Ling, 2010) and McConnaughey (McConnaughey, 1964) to Kulczynski 2 (Holliday et al., 2002) (Table 3). They are directly correlated and are labeled in dark pink (Table 3). The directly correlated coefficients can be used interchangeably in the study of genetic distance of *E. coli*.

The cut-off for Pearson coefficient of distance coefficients is studied. Given the degree of freedom of 1816 (n = 1818), a Pearson's correlation coefficient of 0.115 will be significant at 99.999% confidence based on a previously implementation (Chay & Ling, 2010). This suggest that a Pearson's correlation coefficient of larger than 0.115 will be statistically significant at greater than 99.999% confidence. Hence, the colour coded cells in Table 3 are correlated and can be used interchangeably.

Unexpectedly, Dice (Dice, 1945) is statistically similar to Russel and Rao (Russel & Rao, 1940), and is among the 4 that passed the COV analysis. Russel and Rao can only be used for specific instances due to its exclusion of negative co-occurrences (Coefficient D) in numerator and inclusion in the denominator and it acts as a viable substitute for Dice in this study. Although Dice (Dice, 1940) and Russel and Rao (Russel & Rao, 1945) are different in their mathematical formula, it has a $r^2$ value of 0.998 (Table 3) and had high compactness (Figure 3), suggesting that the both distance coefficients are similar and can be used interchangeably, and be used as a substitute for Dice. The COV values of Russel and Rao (Russel & Rao, 1940) are similar to that of Dice as well (Dice, 1945). Russel and Rao (Russel & Rao, 1940) had correlation value of 0.998, suggesting that the values will differ based on the following formula: $Dice = [Russel\ and\ Rao \pm (1 - r^2)](Russel\ and\ Rao)$. Russel and Rao (Russel & Rao, 1940) can be used to replace Dice (Dice, 1945) it is 99.8% reflective of Dice. Hence, our results suggest that Dice (Dice, 1945) which are used for genetic studies of E. *coli* can be substituted with Russel and Rao (Russel & Rao, 1940).

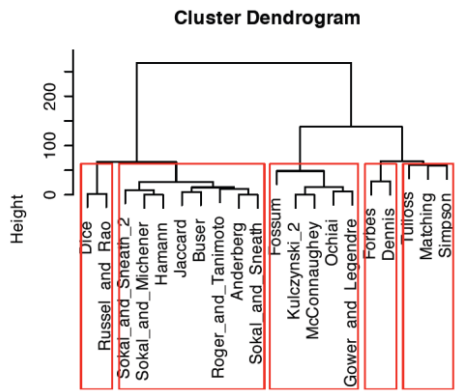

**Figure 3:** Dendrogram of 20 distance correlations to study their similarities and uniqueness against each other. 1: Jaccard; 2: Sokal and Michener; 3: Matching; 4: Dice; 5: Ochiai; 6: Anderberg; 7: Kulczynsk; 8: Forbes; 9: Hamann; 10: Simpson; 11: Russel and Rao; 12: Roger and Tanimoto; 13: Sokal and Sneath; 14: Sokal and Sneath 2; 15: Buser; 16: McConnaughey; 17: Dennis; 18: Gower and Legendre; 19: Tulloss; 20: Fossum.

| | Jaccard | Sokal and Michener | Matching | Dice | Ochiai | Anderberg | Kulczynski 2 | Forbes | Hamann | Simpson | Russel and Rao | Roger and Tanimoto | Sokal and Sneath | Sokal and Sneath 2 | Buser | Fossum | McConnaughey | Dennis | Gower and Legendre | Tulloss |
|---|---|---|---|---|---|---|---|---|---|---|---|---|---|---|---|---|---|---|---|---|
| Jaccard | 1.000 | | | | | | | | | | | | | | | | | | | |
| Sokal and Michener | 0.938 | 1.000 | | | | | | | | | | | | | | | | | | |
| Matching | 0.070 | 0.145 | 1.000 | | | | | | | | | | | | | | | | | |
| Dice | 0.748 | 0.623 | 0.031 | 1.000 | | | | | | | | | | | | | | | | |
| Ochiai | 0.085 | 0.080 | 0.010 | 0.058 | 1.000 | | | | | | | | | | | | | | | |
| Anderberg | 0.969 | 0.082 | 0.068 | 0.722 | 0.082 | 1.000 | | | | | | | | | | | | | | |
| Kulczynski 2 | 0.089 | 0.083 | 0.010 | 0.061 | 0.971 | 0.088 | 1.000 | | | | | | | | | | | | | |
| Forbes | 0.000 | 0.000 | 0.094 | 0.014 | 0.002 | 0.000 | 0.004 | 1.000 | | | | | | | | | | | | |
| Hamann | 0.938 | 1.000 | 0.145 | 0.623 | 0.080 | 0.907 | 0.083 | 0.000 | 1.000 | | | | | | | | | | | |
| Simpson | 0.013 | 0.012 | 0.002 | 0.008 | 0.188 | 0.011 | 0.301 | 0.016 | 0.012 | 1.000 | | | | | | | | | | |
| Russel and Rao | 0.749 | 0.623 | 0.031 | 0.998 | 0.058 | 0.723 | 0.061 | 0.015 | 0.623 | 0.008 | 1.000 | | | | | | | | | |
| Roger and Tanimoto | 0.932 | 0.968 | 0.133 | 0.621 | 0.079 | 0.958 | 0.085 | 0.000 | 0.968 | 0.011 | 0.621 | 1.000 | | | | | | | | |
| Sokal and Sneath | 0.969 | 0.907 | 0.068 | 0.722 | 0.082 | 1.000 | 0.088 | 0.000 | 0.907 | 0.011 | 0.723 | 0.958 | 1.000 | | | | | | | |
| Sokal and Sneath 2 | 0.889 | 0.972 | 0.149 | 0.589 | 0.076 | 0.813 | 0.077 | 0.001 | 0.972 | 0.012 | 0.589 | 0.886 | 0.813 | 1.000 | | | | | | |
| Buser | 0.981 | 0.976 | 0.103 | 0.700 | 0.083 | 0.934 | 0.087 | 0.000 | 0.976 | 0.015 | 0.701 | 0.943 | 0.934 | 0.951 | 1.000 | | | | | |
| Fossum | 0.045 | 0.040 | 0.001 | 0.041 | 0.494 | 0.046 | 0.474 | 0.020 | 0.040 | 0.051 | 0.040 | 0.042 | 0.046 | 0.036 | 0.040 | 1.000 | | | | |
| McConnaughey | 0.089 | 0.083 | 0.010 | 0.061 | 0.971 | 0.088 | 1.000 | 0.004 | 0.083 | 0.301 | 0.061 | 0.085 | 0.088 | 0.077 | 0.087 | 0.474 | 1.000 | | | |
| Dennis | 0.000 | 0.001 | 0.041 | 0.010 | 0.018 | 0.000 | 0.023 | 0.645 | 0.001 | 0.048 | 0.010 | 0.000 | 0.000 | 0.002 | 0.000 | 0.001 | 0.023 | 1.000 | | |
| Gower and Legendre | 0.087 | 0.082 | 0.010 | 0.060 | 0.976 | 0.085 | 0.918 | 0.001 | 0.082 | 0.087 | 0.060 | 0.083 | 0.085 | 0.077 | 0.084 | 0.506 | 0.918 | 0.011 | 1.000 | |
| Tulloss | 0.070 | 0.065 | 0.004 | 0.063 | 0.360 | 0.065 | 0.347 | 0.000 | 0.065 | 0.030 | 0.063 | 0.063 | 0.065 | 0.061 | 0.067 | 0.183 | 0.347 | 0.001 | 0.375 | 1.000 |

**Table 3:** Coefficient of Determination table ($r^2$) after comparisons between 2 tests. $r^2$ values are colour coded based on the intervals. Cells left uncoloured falls below 0.25 interval.

Legend:
- Directly Correlated
- 0.95 – 1
- 0.90 – 0.95
- 0.85 – 0.90
- 0.80 – 0.85
- 0.75 – 0.80
- 0.70 – 0.75
- 0.50 – 0.70
- 0.25 – 0.50

**References**

Basilevsky, A. (1983) *Applied matrix algebra in the statistical sciences*, Dover Publications.

Cavalli-Sforza, L. L. & Bodmer, W. F. (1971) *The genetics of Human populations,* San Francisco, W. H. Freeman and Company.

Bourque, R., Lallier, R., Lariviere, S. (1980) The influence of oral antibiotics on resistance and enterotoxigenicity of *Escherichia coli*. *Canadian Journal of Comparative Medicine,* 44 (1), 101 - 108.

Chay, Z. E., Ling, M. H. T. (2010) COPADS, II: Chi-Square, F-test and t-test Routines from Gopal Kanji's 100 Statistical Tests.

Clermont, O., Johnson, J. R., Menard, M., Denamur, E. (2007) Determination of *Escherichia coli* O types by allele-specific polymerase chain reaction: application to the O types involved in human septicemia. *Diagnostic Microbiology and Infectious Disease*, 57, 129 - 136.

Dice, L. R. (1945) Measures of the amount of ecologic association between species. *Ecology,* 26**,** 6.

Dennis, S. F. (1965) The construction of a thesaurus automatically from a sample of text. In: Stevens, ME, Guiliano, VE and Heilprin, LB (eds). Statistical association techniques for mechanized documentation: Symposium proceedings. National Bureau of Standards. Miscellaneous publication 269.

Dunn, G, Everitt, B. S. (1982). An introduction to mathematical taxonomy. Dover Publications.

Duarte, J. M., Santos, J. B. D. & Melo, L. C. (1999) Comparison of similarity coefficients based on RAPD markers in the common bean. *Genetic Molecular Biology,* 22**,** 6.

Evett, E. W., Scranage, J. & Pinchin, R. (1993) An illustration of the Advantages of Efficient Statistical Methods for RFLP Analysis in Forensic Science. *Genetics,* 52**,** 8.

Fallaw, W. C. (1979) A test of the Simpson coefficient and other binary coefficients of faunal similarity. *Journal of Paleontology,* 53,4, 1029.

Foley, S. L., Lynne, A. M., Nayak, R. (2009) Molecular typing Methodologies for Microbial Source Tracking and Epidemiological investigations of Gram-negative Bacterial Foodborne Pathogens. *Infection, Genetics and Evolution*, 9, 430-440.

Forbes, S. A. (1907). On the local distribution of certain Illinois fishes. Bulletin of the Illinosis State Laboratory of Natural History 7:8.

Fossum, E. G. (1966). Optimization and standardization of information retrieval language and system. Springfield VA: Clearinghouse for Federal Scientific and Technical Information.

Gilbride, K. A., Lee, D.-Y. & Beaudette, L. A. (2006) Molecular Techniques in Wastewater: Understanding microbial communities, detecting pathogens, and real-time process control. *Journal of Microbiological Techniques,* 66**,** 20.

Goldstein, D. B., Linares, A. R., Cavalli-Sforza, L. L. & Feldman, W. W. (1995a) Genetic Absolute Dating Based on Microsatellites and the Origin of Modern Humans. *Proceedings of the National Academy of Sciences USA,* 92**,** 5.

Goldstein, D. B., Linares, A. R., Cavalli-Sforza, L. L. & Feldman, M. W. (1995b) An Evaluation of Genetic Distances for Use with Microsatellite Loci. *Genetics Society of America,* 139.

Gower, J. C., Legendre, P. (1986). Metric and Euclidean properties of dissimilarity coefficients. Journal of Classificatiion 5:5-48.

Hwang, S.A., Yang B.Z., Fitzgerald, E. F., Bush, B., Cook, K. (2001) Fingerprinting PCB patterns among Mohawk women. *Journal of Exposure Science and Environmental Epidemiology,* 11, 184 - 192.

Hamann, U. (1961) Error detecting and error correcting codes. Bell System Technical Journal 29(2): 147 - 160.

Holliday, K. D., Hu, C. Y., Willett, P. (2002). Grouping of coefficients for the calculation of inter-molecular similarity and dissimilarity using 2D fragment bit-strings. Combinatorial Chemistry & High Throughput Screening 5(2): 155-166.

Jaccard, P. (1908) Nouvelles recherches sur la distribution florale. *Bull Soc Vaud Sci Nat,* 44, 223-270.

Li, W.-M. (1981) A Simulation study on Nei and Li's model for estimating DNA divergence rom restriction enzyme maps. *Journal of Molecular Evolution*, 17, 4.

Ling, M. (2010) Distances Coefficients between Two Lists or Sets. *The Python Papers Source Codes,* 2, 2.

Mcconnaughey, H. H. (1964) The determination and analysis of plankton communities. Penelitian laut de Indonesia (Marine research in Indonesia), Special number, 1-40.

MacDonald, K. L., O'Leary, M. J., Cohen, M. L., Norris, P., Wells, J. G., Noll, E., Kobayashi, J. M., Blake, P. A. (1988) Escherichia coli 0157:H7, an Emerging Gastrointestinal Pathogen. *The Journal of the American Medical Association,* 259, 24, 3567 - 3570.

Meyer, A. D. S., Garcia, A. A. F., Souza, A. P. D., Souza Jr, C. L. D. (2004) Comparison of similarity coefficients used for cluster analysis with dominant markers in maize (*Zea Mays* L). *Genetics and Molecular Biology,* 27,

Nei, M. & Li, W.-H. (1979) Mathematical model for studying genetic variation in terms of restriction endonucleases. *Genetics,* 76**,** 5.

Ochiai, A. (1957). Zoogeographic studies on the soleoid fishes found in Japan and its neighbouring regions. *Bull. Jpn. Soc. Sci. Fish,*22, 526 -530.

O' Fallon, E., Pop-Vicas, A., D`Agata, E. (2009) The Emerging Threat of Multidrug-resistant gram-negative organisms in long-term care facilities. *Journals of Gerontology Series A, Biological Sciences and Medical Sciences,* 62 (1), 138 - 141.

Reynolds, J., Weir, B. S. & Cockerham, C. C. (1983) Estimation of the Coancestry Coefficient: Basis for a Short-Term Genetic Distance. *Genetics Society of America,* 105**,** 13.

Rogers, D. J. and Tanimoto, T. T. (1960). A computer program for classifying plants. *Science,* 132, 1115-1118.

Russel, P. F., Rao, T. R. (1940) On habitat and association of species of anopheline larvae in south-eastern Madras. J. Malaria Inst. India 3: 154-178.

Shriver, M. D., Jin, L., Boerwinkle, E., Deka, R., E. Ferrell, R. & Chakraborty, R. (1995) A Novel Measure of Genetic Distance for Highly Polymorphic Tandem Repeat Loci. *Molecular Biology Evolution,* 12**,** 7.

Silva, M.F., Vaz-Moreira, I., Gonzalez-Pajuelo, M., Nunes, O. C., Manaia, C. M. (2007) Antimicrobial resistance patterns in *Enterobacteriaceae* isolated from an urban wastewater treatment plant. *Federation of European Microbiological Societies Microbiology Ecology,* 60, 166-176.

Sokal, R. R., Michener, C. D. (1985). A statistical method for evaluating systematic relationships. Univ Kansas Sci Bull 38:1409 - 1438.

Sokal, R. R, Sneath, P. H. A. (1963) Principles of Numeric Taxonomy. W. H. Freeman, San Francisco.

Tulloss, R. E. (1997) Assessment of similarity indices for undesirable properties and a new tripartite similarity index based on cost functions. In: Palm, M. E. and I. H. Chapela, eds. Mycology in Sustainable Development: Expanding Concepts, Vanishing Borders. Parkway Publishers, Boone, North Carolina.

Wang, R., Zheng, L., Toure, Y. T., Dandekar, T., Kafatos, C. F. (2001) When genetic distance matters: Measuring genetic differentiation at microsatellite loci in whole genome scans of recent and incipient mosquito species. *Proceedings of the National Academy of Sciences,* 98**,** 6.

Welch, R. A., Burland, V., Plunkett III, G., Redford, P., Roesch, P., Rasko, D., Buckles, E. L., Liou, S.-R., Boutin, A., Hackett, J., Stroud, D., Mayhew, G. F., Rose, D. J., Zhou, S., Schwartz, D. C., Perna, N. T., Moblet, H. L. T., Donnenberg, M. S., Blattner, F. R. (2002) Extensive mosaic structure revealed by the complete genome sequence of uropathogenic *Escherichia coli.Proceedings of the National Academy of Sciences*, 99, 26, 17020 - 17024.

## Appendix

```
R Code for Hierarchical Clustering

data = read.table('Data.csv', header=TRUE, sep=',')
data = na.omit(data)
data = scale(data)
data = t(data)
d = dist(data, method='euclidean')
fit = hclust(d, method='ward')
plot(fit)
g = cutree(fit, k=5)
rect.hclust(fit, k=5, border='red')
```